\documentclass[journal]{IEEEtran}
\usepackage{verbatim}
\usepackage{amsfonts}

\usepackage{graphicx}
\usepackage{amsmath,amssymb} 
\usepackage{color}
\usepackage{epstopdf}
\usepackage{pifont}
\usepackage{booktabs}
\usepackage{makecell}

\usepackage{tabularx}
\usepackage{float}
\usepackage[colorlinks,linkcolor=black]{hyperref}
\usepackage{multirow}
\usepackage{makecell}
\usepackage{color}
\usepackage{graphicx}
\ifCLASSINFOpdf
\else
\fi

\begin{document}
\title{Illumination Histogram Consistency Metric for Quantitative Assessment of Video Sequences}
\author{
Long~Chen,
Mobarakol~Islam,
Thomas~Dowrick
\thanks{Long Chen and Mobarakol Islam are with Department of Medical Physics and Biomedical Engineering, University College London, United Kingdom, e-mail: (chenlongcv@gmail.com).}
\thanks{Thomas Dowrick is with x, United Kingdom, e-mail: (xxx@gmail.com). Thomas is the corresponding author. }
\thanks{Manuscript received July 1, 2020; revised xxxxx.}}

\markboth{}
{Shell \MakeLowercase{\textit{et al.}}: Bare Demo of IEEEtran.cls for IEEE Journals}

\maketitle
\begin{abstract}

The advances in deep generative models have greatly accelerate the process of video procession such as video enhancement and synthesis. Learning spatio-temporal video models requires to capture the temporal dynamics of a scene, in addition to the visual appearance of individual frames. Illumination consistency, which reflects the variations of illumination in the dynamic video sequences, play a vital role in video processing. Unfortunately, to date, no well-accepted quantitative metric has been proposed for video illumination consistency evaluation. In this paper, we propose a illumination histogram consistency (IHC) metric to quantitatively and automatically evaluate the illumination consistency of the video sequences. IHC measures the illumination variation of any video sequence based on the illumination histogram discrepancies across all the frames in the video sequence. Specifically, given a video sequence, we first estimate the illumination map of each individual frame using the Retinex model; Then, using the illumination maps, the mean illumination histogram of the video sequence is computed by the mean operation across all the frames; Next, we compute the illumination histogram discrepancy between each individual frame and the mean illumination histogram and sum up all the illumination histogram discrepancies to represent the illumination variations of the video sequence. Finally, we obtain the IHC score from the illumination histogram discrepancies via normalization and subtraction operations. Experiments are conducted to illustrate the performance of the proposed IHC metric and its capability to measure the illumination variations in video sequences. The source code is available on \url{https://github.com/LongChenCV/IHC-Metric}.

\end{abstract}
%
\IEEEpeerreviewmaketitle
\section{Introduction}

Deep generative models have achieved remarkable success in various video translation tasks \cite{chen2019mocycle, wei2018video} such as video synthesis \cite{liu2021generative, li2022neural} and video enhancement \cite{peng2022lve, andrei2021supervegan}, in which the deep generative model not only needs to capture the realistic visual distributions but also the temporal dynamics of the target video domain. Video quality evaluation is the foundation of video translation tasks as it provides reliable guidance in video generation. A few studies have advanced the research field of video quality evaluation \cite{hou2020no, tu2021ugc}, such as realism evaluation and temporal consistency evaluation \cite{bansal2018recycle, rivoir2021long}, however, no work has proposed illumination consistency metric to evaluate the illumination variations in video translations.

Effective and objective evaluation metric for illumination consistency measure is a critical component in many video translation applications, such as novel view synthesis~\cite{gu2023nerfdiff, tretschk2021non} and missing frame prediction~\cite{chaubey2023estimation, bao2019memc}. However, to date, no specific metric is designed for the evaluation of illumination consistency and only human subjective evaluation is used to evaluate the illumination consistency. Nevertheless, the subjective evaluation requires human interventions and is usually biased. Therefore, it is important to have a reliable objective evaluation metric to quantitatively and automatically measure the illumination consistency. 

In this work, we aim at presenting a simple and objective metric for quantitatively assessing the illumination consistency of video sequences. In order to quantitatively measure the illumination consistency of video sequence, we propose a novel Illumination Histogram Consistency (IHC) metric, which is developed based on the illumination histogram extracted from the Retinex model~\cite{hou2017image, hussein2019retinex}. IHC is a simple and fast solution for illumination consistency evaluation in video sequences. More importantly, it can also applied for illumination consistency evaluation in any dispersed image datasets in addition to consecutive video sequences.


\begin{figure*}[tb]
\centering
\includegraphics[width=18cm]{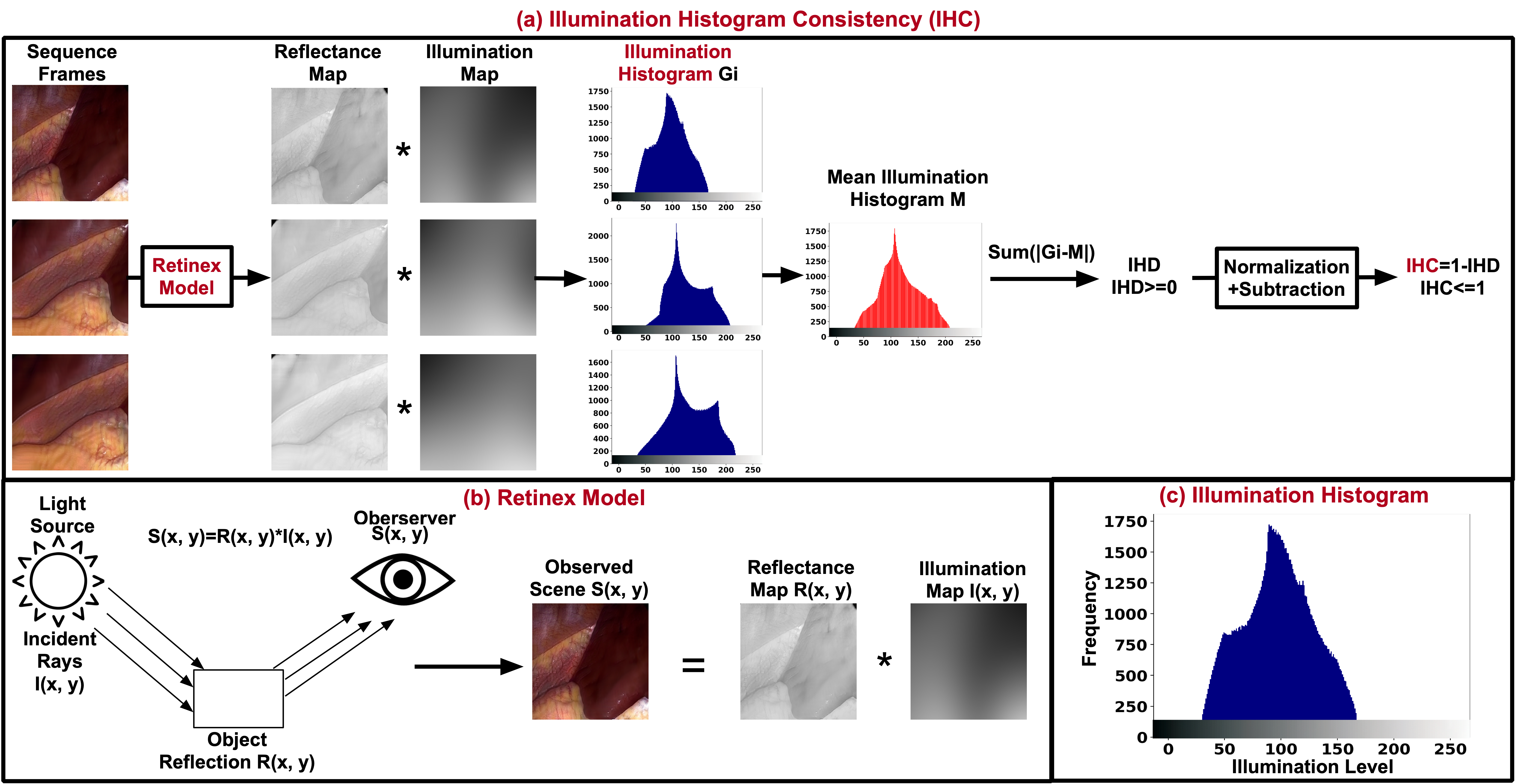}
\caption{The proposed illumination histogram consistency (IHC) metric (a). For the video sequence, IHC first estimates the illumination map using the Retinex Model (b). Then, it exploits the consistency of illumination histogram (c) to measure the illumination consistency of the video sequence.}
\label{fig:fiducial}
\end{figure*}

\section{The proposed IHC Metric}

In this work, we aim to present a quantitative metric to evaluate the illumination consistency of video sequences. To evaluate the illumination consistency in a video sequence, comparing the illumination components is more straightforward and intuitive than comparing the raw images. Hence, we first estimate the illumination map of each raw image using the Retinex model. Then, we extract the illumination component of individual image from the illumination map. We exploit illumination histogram to quantitatively measure the illumination component of each image. The illumination histogram presents the number of pixels for each brightness level (from black to white), intuitively reflecting the illumination distribution of individual image. Next, we compute the illumination histogram consistency (IHC) metric of the whole video sequence. Intuitively, the IHC metric must capture the illumination variations of the video sequence, in this work, the illumination variations of the video sequence is quantitatively measured by the variations of illumination histogram across all the frames in the video sequence. 

\begin{figure*}[tb]
\centering
\includegraphics[width=18cm]{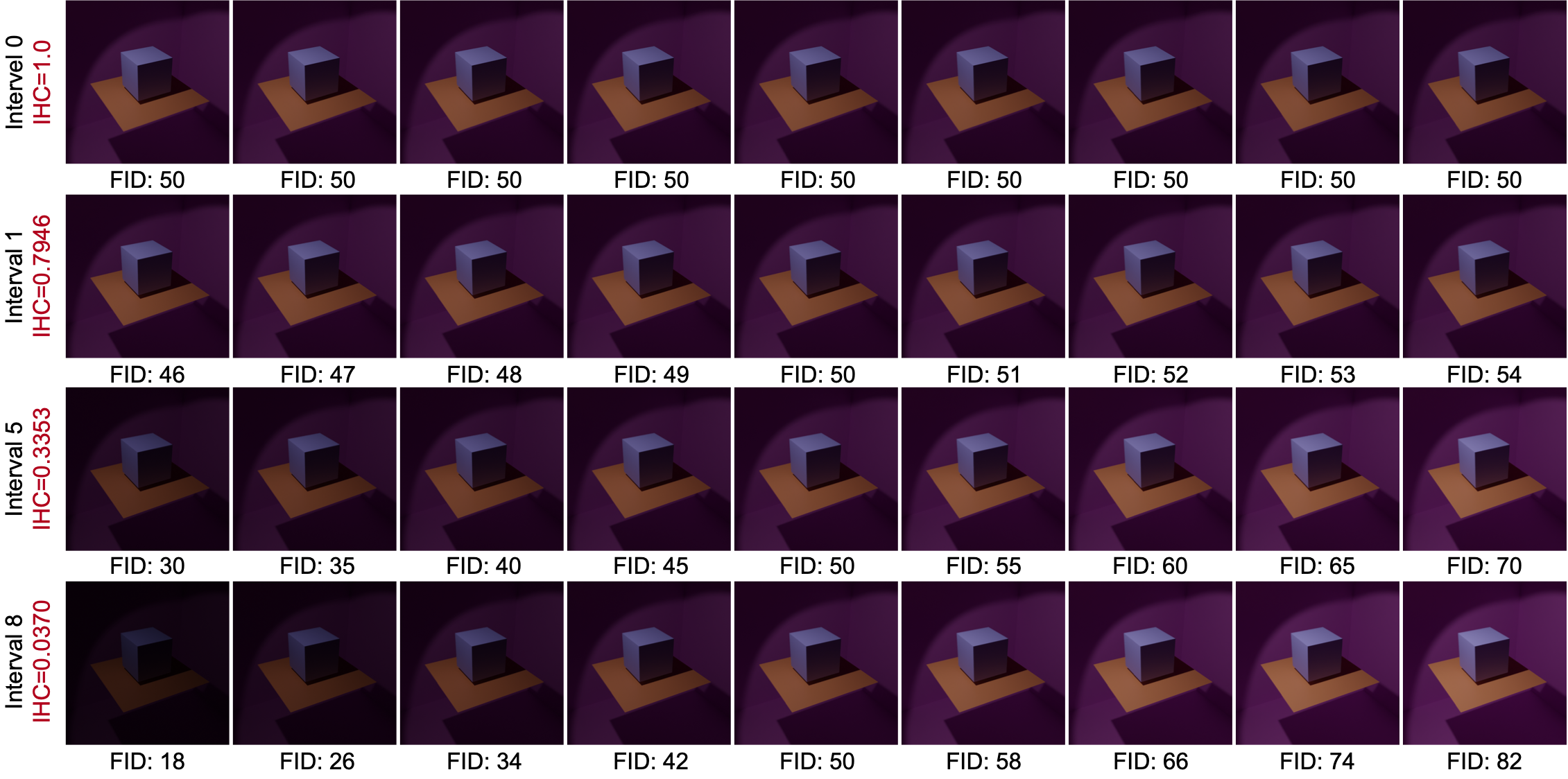}
\caption{The IHC scores of image sets generated by different frame interval settings. 'FID' indicates the frame ID in the video sequence with linearly increasing illumination density. 'Interval' indicates the interval between two adjacent frames.}
\label{fig:IHC_intervals}
\end{figure*}

\begin{figure*}[tb]
\centering
\includegraphics[width=18cm]{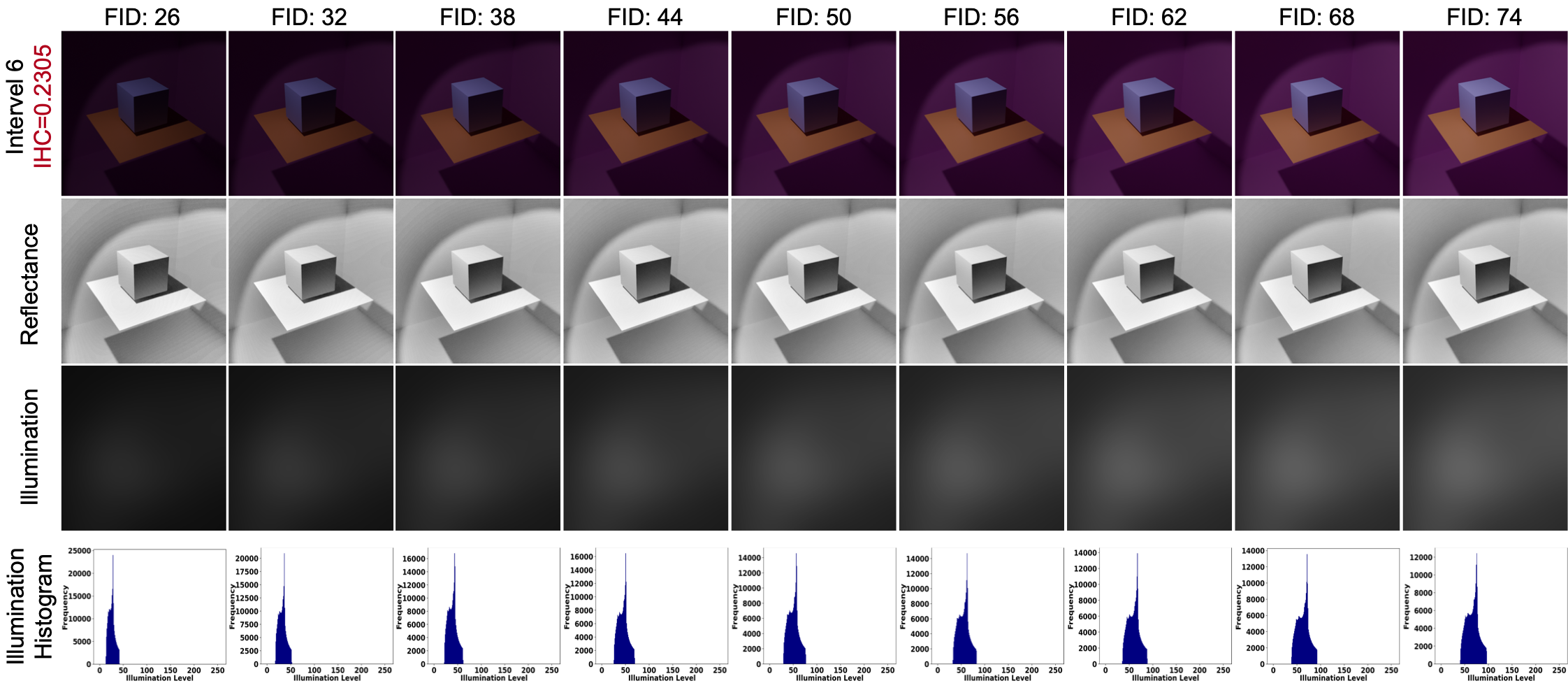}
\caption{The raw images, the reflectance maps, the illumination maps and the illumination histograms generated under the setting of $Interval=6$.}
\label{fig:illumination_reflection}
\end{figure*}
\subsection{Illumination Map Estimation}

In the proposed IHC framework, the first step is to estimate the illumination map of individual frame in the video sequence. We derive the illumination map using the single-scale Retinex model~\cite{bogdanova2010image, hines2005single}, which aims to simulate the Retina and Cortex units in human visual system. Retinex model describes an observed image scene as the product of the illumination and reflectance of the objects in the scene and can be formulated as Eq.~\ref{eq:scene}:
\begin{equation}
S(x, y) = R(x, y)*I(x, y)
\label{eq:scene}
\end{equation}
Where $S$ is the image scene perceived by human visual system. $R$ is the reflectance map of the object surface and its range lies between 0 and 1, $L$ is the illumination map on the object surface and its range lies between the value  0 and 255. The characteristics of the illumination map depend on the source of illumination and the characteristics of the reflectance map depend on the nature of object. 

Mathematically, we can obtain illumination map by dividing the image with the reflectance map, however, it is impractical to obtain reflectance without any assumption. To estimate the illumination and reflectance maps, different assumptions have been constructed. Single-scale Retinex model is one of the most commonly used models to estimate the reflectance map $R$, which can be formulated as: 

\begin{equation}
R(x, y) = log S(x, y)-log[F(x, y) * S(x, y)]
\label{eq:retinex}
\end{equation}
where $S(x, y)$ is the image scene perceived by human visual system at pixel position $(x, y)$, $R(x, y)$ is the reflectance of image at pixel position $(x, y)$, $*$ denotes the convolution operation and $F(x, y)$ is a Gaussian function, which can be expressed as 
\begin{equation}
F(x, y) = \frac{1}{\sqrt{2\pi\sigma}}e^{-\frac{x^2+y^2}{2\sigma^2}}
\label{eq:gaussian}
\end{equation}

Usually, the Gaussian function $F(x, y)$ is empirically selected, then, the reflectance map $R$ and the illumination map $I$ can be obtained using the single-scale Retinex model. 

\subsection{Illumination Histogram Consistency (IHC) Score}

The IHC score is calculated from the illumination maps across all the frames in the video sequence, the computation process can be formulated as follows
\begin{itemize}

\item Given a video sequence containing $K$ image frames, we use the Retinex model to extract the illumination map for each frame in the video sequence.

\item Obtain the illumination histogram $G_i$ of each image from the illumination map. $G_i\ (\ 0\leq i \leq K)$ denotes the illumination histogram of the $i$-th frame.

\item Compute the mean illumination histogram $M$ of all the frames using the mean operation.

\begin{equation}
M(j) = \frac{\sum_{i=1}^{K}|G_i(j)|}{K}, 0<=j<=255
\label{eq:CE}
\end{equation}
$G_i(j)$ indicate the number of the pixels whose values are equal to $j$ in $G_i$, the illumination/brightness has been divided into 255 levels.

\item For each image, compute the difference between its illumination histogram and the mean illumination histogram $|G_i-M|$.

\item Sum the illumination discrepancies of all images and use the summed illumination discrepancy to represent the illumination changes of the whole sequence.

\item Normalize the illumination histogram discrepancy $IHD$.
\begin{equation}
IHD = \frac{\sum_{i=1}^{K}\sum_{j=0}^{255}|G_i(j)-M(j)|}{K*S}
\label{eq:CE}
\end{equation}

\item Compute the illumination histogram consistency $IHC$ from the illumination histogram discrepancy $IHD$ by the subtraction operation.

\end{itemize}

\begin{equation}
IHC =2- \frac{\sum_{i=1}^{K}\sum_{j=0}^{255}|G_i(j)-M(j)|}{K*S}
\label{eq:CE}
\end{equation}

where $K$ indicates the number of images in the image set to be evaluated. $G_i$ indicates the illumination histogram of the $i$-th image, and $G_i(j)$ indicate the number of the pixels whose values are equal to $j\ (0<=j<=255)$ in $G_i$. $M(j)$ is the mean histogram computed over the whole image set and $S$ is the size of the image, i.e. the pixel number of the images. For example, for an image of width $w$ and height $h$, its size $S=w*h$.

\section{Experiments}

\begin{figure}[tb]
\centering
\includegraphics[width=8cm]{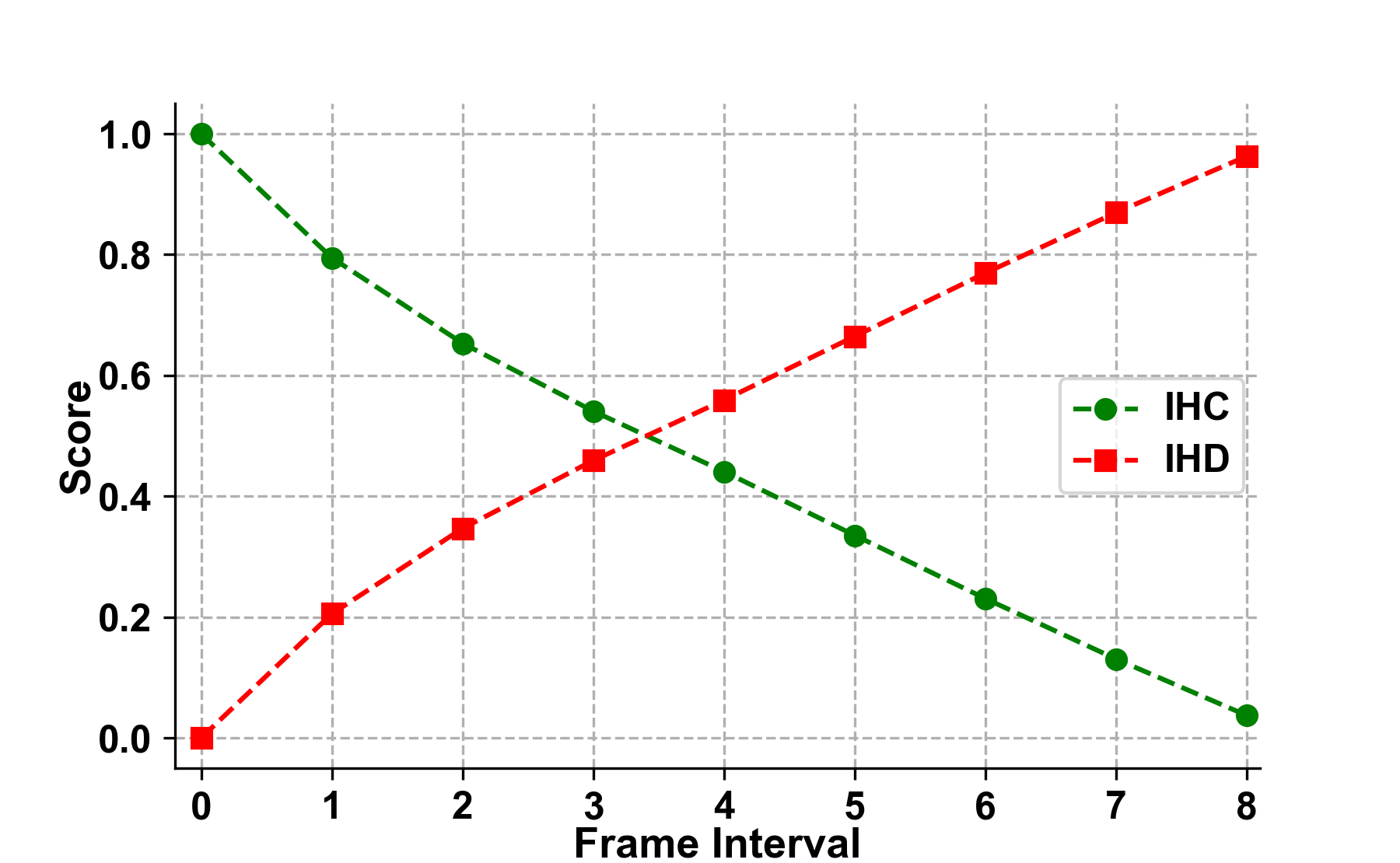}
\caption{Larger frame intervals bring higher IHD score and lower IHC scores since more illumination variations are introduced by larger intervals.}
\label{fig:intervals}
\end{figure}

\subsection{Sequential Data Generation}

To valid the effectiveness of the proposed IHC metric, we exploit the 3D software Blender \url{https://www.blender.org/} to generate 100 continuous video frames with linearly increasing illumination. The simulated video frames can be downloaded on \url{https://github.com/LongChenCV/IHC}. 

Taking the 50-th frame as the center, we adopt different, fixed intervals to extract 4 frames from the front and back, forming multiple image sets under different interval settings. Some examples of the image sets are presented in Fig.~\ref{fig:IHC_intervals}. 

\subsection{Results and Discussions}

Fig.~\ref{fig:illumination_reflection} presents the raw images, the reflectance maps, the illumination maps and the illumination histograms generated under the setting of $Interval=6$. The reflectance map and illumination map are estimated by the Retinex model from the raw image, and the illumination histogram is extracted from the illumination map and used to compute the final IHC score. From the observation of the illumination map, we perceive evident, increasing illuminations from the left to the right. 

Theoretically, large intervals bring smaller IHC scores as they introduce more illumination variations. To verify this theoretical assumption, we compute the IHC and IHD scores under the settings of differnt intervals, and present the results in Fig.~\ref{fig:intervals}, from which we observe that the IHC scores increase with the increasing intervals nearly linearly. This demonstrate our IHC metric can capture the illumination variations in the video sequence. The experimental results demonstrate that IHD effectively evaluates the illumination discrepancies and IHC effectively evaluate the illumination consistency in accordance with the human perceptions.

\section{Conclusions}

We proposed the Illumination Histogram Consistency, a new evaluation metric for illumination consistency of video sequences, and an important step towards better evaluation of deep generative models for video translations. Our experiments confirm that IHC is accurate in evaluating video illumination consistency. More importantly, IHC can be applied for illumination consistency evaluation in any dispersed image sets in addition to consecutive video sequence.

\bibliographystyle{IEEEtran}
\bibliography{mybibfile}

\end{document}